\documentclass[lettersize,journal]{IEEEtran}

\usepackage{algorithm}

\usepackage{textcomp}
\usepackage{stfloats}
\usepackage{url}
\usepackage{verbatim}
\usepackage{graphicx}
\usepackage{cite}

\usepackage{amsmath,amssymb,amsfonts}
\usepackage{algorithmic}
\usepackage{cite}
\usepackage{booktabs} 
\usepackage{amsmath,amssymb,amsfonts}
\usepackage{algorithmic}
\usepackage{multirow}
\usepackage{multirow}
\usepackage{booktabs}
\usepackage{geometry}

\usepackage{graphicx}
\usepackage{lipsum}
\usepackage{float}
\usepackage{subcaption} 
\usepackage{caption} 
\usepackage{textcomp}
\def\BibTeX{{\rm B\kern-.05em{\sc i\kern-.025em b}\kern-.08em
		T\kern-.1667em\lower.7ex\hbox{E}\kern-.125emX}}
\usepackage{balance}

\begin{document}

\title{A Novel Framework using Intuitionistic Fuzzy Logic with U-Net and U-Net++ Architecture: A case Study of MRI Bain Image Segmentation}

\author{Hanuman Verma,Kiho Im, Akshansh Gupta and M. Tanveer
	\thanks{This paragraph of the first footnote will contain the date on which
		you submitted your paper for review. It will also contain support information,
		including sponsor and financial support acknowledgment. For example, 
		``This work was supported in part by the U.S. Department of Commerce under Grant BS123456.'' }
	\thanks{Hanuman Verma is with the Department of Mathematics, Bareilly College, Bareilly (MJP Rohilkhand University), Uttar Pradesh, 243005, India (e-mail: hv4231@gmail.com). }
	\thanks{Kiho Im is with 2Fetal Neonatal Neuroimaging and Developmental Science Center, Boston Children’s Hospital, Harvard Medical School, Boston, MA 02115, USA
		and Division of Newborn Medicine, Boston Children’s Hospital, Harvard Medical School, Boston, MA 02115, USA
		also with Department of Pediatrics, Harvard Medical School, Boston, MA, USA
		(e-mail:Kiho.Im@childrens.harvard.edu).}
	\thanks{Akshansh Gupta is with National Institute of Science Communication and Policy Research, New Delhi, 110012, India (e-mail: akshanshgupta83@gmail.com).}
	\thanks{Prof M. Tanveer is with the Department of Mathematics, Indian Institute of Technology Indore, Indore (M.P.)-453552, India  (e-mail: mtanveer@iiti.ac.in).}

}

\maketitle

\begin{abstract}
	Accurate segmentation of brain images from magnetic resonance imaging (MRI) scans plays a pivotal role in brain image analysis and the diagnosis of neurological disorders. Deep learning algorithms, particularly U-Net and U-Net++, are widely used for image segmentation. However, it finds difficult to deal with uncertainty in images. To address this challenge, this work inte-grates intuitionistic fuzzy logic into U-Net and U-Net++, propose a novel framework, named as \text{IFS\_U-Net} and \text{IFS\_U-Net++}. These models accept input data in an intuitionistic fuzzy repre-sentation to manage uncertainty arising from vagueness and imprecise data. This approach ef-fectively handles tissue ambiguity caused by the partial volume effect and boundary uncertain-ties. To evaluate the effectiveness of \text{IFS\_U-Net} and \text{IFS\_U-Net++}, experiments are conducted on two publicly available MRI brain datasets: the Internet Brain Segmentation Repository (IBSR) and the Open Access Series of Imaging Studies (OASIS). Segmentation performance is quantitatively assessed using Accuracy, Dice Coefficient, and Intersection over Union (IoU). The results demonstrate that the proposed architectures consistently improve segmentation per-formance by effectively addressing uncertainty.
\end{abstract}

\begin{IEEEkeywords}
	U-Net, U-Net++, Intuitionistic fuzzy set, MRI, Brain image segmentation.
\end{IEEEkeywords}

\section{Introduction}
\label{introduction}
Accurate segmentation of brain images from magnetic resonance imaging (MRI) scans plays an essential role in brain image analysis and acts as a key player in neurological disease diagnosis and treatment planning \cite{Despotovic2015}. The brain image segmentation is a process to partition the brain MRI images into various tissues: CSF (Cerebrospinal Fluid), GM (Gray Matter), and WM (White Matter), and it is used for detecting and analyzing neurological disorders disease such as tumors, multiple sclerosis, Alzheimer's disease, stroke, etc. The manual segmentation of brain images from MRI is more reliable, but it is time-consuming and prone to inconsistencies among experts. As a result, automated segmentation techniques come in the mind.  Many automatic segmentation techniques based on unsupervised and supervised machine learning algorithms have been developed to segment MRI brain images \cite{Verma2016,Gupta2023,Fawzi2021,Wang2022}. 

Recently, deep learning techniques, mainly convolutional neural networks (CNNs) based architecture, a subset of machine learning algorithms, have gotten significant attention for brain image segmentation \cite{Moeskops2016} . Further, advancements in convolutional neural networks, U-Net (Ronneberger et al., 2015) and several modified versions of U-Net architecture such as: Attention U-Net \cite{Oktay2018}, Residual U-Net \cite{Zhang2018}, and 3D U-Net \cite{Cicek2016}  have been proposed by researcher to enhance feature extraction, localization capabilities and overall to achieve the segmentation performance. The U-Net architecture is proposed by Ronneberger et al.\cite{Ronneberger}, and it consists of a contracting path for capturing contextual information and an expansive path for precise localization of the features. Further several modified versions of U-Net, have been introduced to improve the image segmentation performance \cite{Azad2024,Punn,Liu2020}. Çiçek et al. \cite{Cicek2016} proposed a 3D U-Net model that extends 2D operations in U-Net to volumetric 3D operation in U-Net, for improving segmentation performance of image. Moreover, Hatamizadeh et al. \cite{Hatamizadeh2021} proposed Swin UNETR, incorporating a hierarchical transformer-based structure and U-Net architecture, which provides the better performance than U-Net and further tested for tumors segmentation in MRI brain images. An optimized U-Net architecture \cite{Futrega2021} has been introduced for brain tumor segmentation task and presented in BraTS21 dataset challenge with using Focal loss, deep supervision loss, decoder attention, drop block, and residual connections in the architectures. Ali et al. \cite{Ali2022} implemented the modified U-Net which reduce the number of parameters and optimize the model for required computational resources and further tested on medical image segmentation. The transformer-based architectures, named TransUNet \cite{Chen2021} , has been introduced with combining the merit of transformers \cite{Vaswani2017} and U-Net architecture. 

Further, U-Net++ architecture \cite{Zhou2018} is designed to address limitations in U-Net by incorporating a series of nested, dense skip connections between the encoder and decoder. These redesigned skip pathways aim to reduce the semantic gap between feature maps of encoder and decoder sub-networks, facilitating more precise segmentation of medical images. Micallef et al. \cite{Micallef2021} explored U-Net++ model for automatic brain tumor segmentation. In this U-Net++ architecture, U-Net is adopted and incorporated modifications in loss function, convolutional blocks, and deep supervision and further tested on biomedical image segmentation, specifically demonstrated on brain tumor segmentation.

In human brain image, uncertainty and vagueness arise due to intensity variations and transition regions in various brain tissue, making it tedious work to segment the pixels into correct brain tissue. Various kind artifact occurs during scan of MRI brain images \cite{Henkelman1985}. One artifact in brain images is the partial volume effect, where a single voxel overlap of multiple brain tissue, leading to ambiguous intensity values. This effect is common at tissue boundaries in the brain image and reduces segmentation accuracy.

Intuitionistic fuzzy logic, based on intuitionistic fuzzy set theory (IFS) \cite{Atanassov1986,Atanassov1989}, enhanced the decision making of fuzzy logic, which is derived from set theory (Zadeh, 1965) by introducing an additional degree of uncertainty, called hesitation, to address imprecise and ambiguous information. In contrast to fuzzy logic, which considers membership (belongingness) and non-membership (non-belongingness), intuitionistic fuzzy logic incorporates a degree of hesitation that addresses ambiguity. This improved paradigm facilitates better decision-making in situations where uncertainty is present in data. It has been extensively utilized in image segmentation, pattern recognition, and decision-making systems where uncertainty is a crucial issue. In medical image computing, intuitionistic fuzzy logic enhances segmentation accuracy. One of the widely used unsupervised clustering techniques based on IFS theory is the intuitionistic fuzzy c-means (IFCM). The IFCM and various variants of IFCM are also used in medical image segmentation \cite{Verma2015, Verma2016, Kumar2019, Verma2025, Yang2021}. 

Currently, with the advancements in deep learning architectures for medical image computing, most of the brain image segmentation algorithms rely on convolutional neural networks, particularly U-Net \cite{Ronneberger}   and its various modifications, such as U-Net++ \cite{Zhou2018}. However, the U-Net architecture is unable to handle uncertainty properly caused by vagueness and the partial volume effect in brain images. Some efforts have been made by researcher in literature to integrate fuzzy logic into deep learning to address this uncertainty. The integration of a fuzzy layer in deep learning was introduced by Price et al. \cite{Price2019}, which utilizing the Choquet and Sugeno fuzzy integrals. This approach uses the aggregation properties in deep learning and is applied to semantic segmentation. Sharma et al. \cite{Sharma2019} proposed an innovative pooling layer based on fuzzy logic within the convolutional layer to extract meaningful information. The novel pooling layer leverages fuzzy logic to extract information in two distinct ways and was later tested for image categorization. Ding et al. \cite{Ding2021} proposed a multimodal model integrating deep learning with fuzzy logic, referred to as fuzzy-informed deep learning segmentation (FI-DL-Seg), for infant MRI brain segmentation. This model supports input images in T1 and T2-weighted formats. The FI-DL-Seg model integrates fuzzy and uncertainty layers, volumetric fuzzy pooling, and fuzzy-enabled multiscale learning to address ambiguity in convolutional feature values. These features assist in retrieving information from uncertain and unclear MRI brain images. Huang et al. \cite{Huang2021} introduced the concept of fuzzy logic in CNNs for segmentation of ultrasound breast images. First, the image is augmented using wavelet transform, and then the augmented image has been transformed into the fuzzy domain with various membership functions to address the uncertainty in the images. Badawy et al. \cite{Badawy2021} introduced a deep learning model that integrates fuzzy logic with a CNNs for the semantic segmentation of breast cancers from ultrasound images. In this approach, the preprocessing steps are carried out using fuzzy logic, while the segmentation is accomplished using CNNs. Huang et al.\cite{Huang2022} suggested a method for breast image segmentation from ultrasound images using fuzzification to reduce uncertainty in CNNs. This approach, known as the spatial and channel-wise fuzzy uncertainty reduction network, tries to minimize uncertainty in image segmentation. Subhashini et al. \cite{Subhashini2022} combines the concepts of the fuzzy logic and deep learning model together to make a three-way decision to handle the uncertainty that arises due to vagueness. To incorporate fuzzy logic in U-Net architecture, Chen et al. \cite{Chen2022} developed the target-aware U-Net model with fuzzy skip connections and applied it to pancreas segmentation. The fuzzy skip connection applies fuzzy logic to extract high-level semantic information. Additionally, a target mechanism is integrated into the decoder section of this model to enhance feature representation for the required segmentation. Nan et al. \cite{Nan2023} introduced the fuzzy attention neural network with integrating the attention with fuzzy logic in 3D U-Net architecture. The designed model reduces the uncertainties in airway segmentation from the computerized tomography images and help to detect the discontinuity in the airway. Deep learning model like U-Net and U-Net++ have the problem to handle the uncertainty like partial volume effect and vagueness in proper way, and need to integrate the intuitionistic fuzzy logic, to further improves segmentation accuracy by utilizing these uncertainties more effectively.

In this work, to integrate the intuitionistic fuzzy logic with U-Net and U-Net++ architecture, we propose a novel framework: intuitionistic fuzzy logic with U-Net (\text{IFS\_U-Net}) and intuitionistic fuzzy logic with U-Net++ (\text{IFS\_U-Net++}) with utilizing input data in intuitionistic representation form. The main contribution in this work includes:

\begin{itemize}
	\item It integrates intuitionistic fuzzy logic by transforming the input image into an intuitionistic fuzzy representation form.
	\item It processes the input image using membership, non-membership, and the crucial hesitation function to manage uncertainty.
	\item By leveraging uncertainty in intuitionistic fuzzy logic, the proposed method effectively captures and utilizes uncertainty, enhancing model robustness and segmentation accuracy.
	\item It improves segmentation accuracy by efficiently handling uncertainties.
	
\end{itemize}

To conduct a comprehensive evaluation and efficacy for proposed framework viz \text{IFS\_U-Net} and \text{IFS\_U-Net++}, experiments are carried out on two publicly available MRI brain datasets: Internet Brain Segmentation Repository (IBSR) and the Open Access Series of Imaging Studies (OASIS). The segmentation performance is evaluated using Accuracy, Dice coefficient, and Intersection over Union (IoU) metrices. These segmentation performance metrics demonstrate the consistent efficacy of proposed architectures.

The paper is structured as follows: Section \ref{Intui} discusses intuitionistic fuzzy sets and the intuitionistic fuzzy representation of data. Section \ref{Intfuzz} introduces the intuitionistic fuzzification of images. Section \ref{Pro} details the proposed novel system viz \text{IFS\_U-Net} and \text{IFS\_U-Net++} architecture. Section \ref{Exp} describes the experimental setup, while Section \ref{Resdis} presents the results and their analysis. Finally, Section \ref{con} concludes the study and outlines the potential future research directions.

\section{Intuitionistic fuzzy sets and intuitionistic fuzzy representation of data}\label{Intui}
The intuitionistic fuzzy set (IFS) A with membership degree (belongingness) $\mu_A \colon X \to [0,1]$ and non-membership degree (non-belongingness) $\nu_A \colon X \to [0,1]$ on a set $X$ is defined by Atanassov (\cite{Atanassov1986,Atanassov1989}): 

\begin{equation}
	A = \left\{ \langle x, \mu_A(x), \nu_A(x) \rangle : \forall x \in X \right\}
	\label{eq:1}
\end{equation} 

In IFS, membership and non-membership degree satisfied the condition $0 \leq \mu_A(x) + \nu_A(x) \leq 1$ and hesitation degree (uncertainty) $\pi_A(x)$ is computed as $\pi_A(x) = 1 - \mu_A(x) - \nu_A(x)$ . If the value of hesitation is zero, then either the points strongly belong to the set or do not strongly belong to the set; that is, the IFS becomes a fuzzy set. 

The IFS theory comprises the membership, non-membership and hesitation degree and it requires in construction of IFS. The membership degree is computed using Triangular, Trapezoidal, Gaussian, and Sigmoid membership function etc.  In literature some negation functions are introduced to compute the non-membership degree like Sugano’s and Yager negation function. Using Sugeno’s negation function \cite{Bustince2000}, the non-membership function is defined as $\mu(x) = \frac{1 - \mu_A(x)}{1 + \lambda \mu_A(x)}, \quad \lambda > 0$. The intuitionistic fuzzy set $A$ in \eqref{eq:1} can be rewritten as:

\begin{equation}
	A_{\text{Sugeno}} = \left\{ \left\langle x, \mu_A(x), \nu_A(x) = \frac{1 - \mu_A(x)}{1 + \lambda \mu_A(x)} \right\rangle : \forall x \in X \right\}
	\label{eq:2}
\end{equation} 

Yager (\cite{Yager1979,Yager1980}) negation function $N(\mu(x)) = \left( 1 - ( \mu_A(x))^{\alpha} \right)^{\frac{1}{\alpha}}, \quad 0 < \alpha < 1
$ used to computed the non-membership degree and the IFS $A$ in rewritten as:

\begin{equation}
	A_{\text{Yager}} = \left\{ \left\langle x, \mu_A(x), \nu_A(x) = \left( 1 - (\mu_A(x))^{\alpha} \right)^{\frac{1}{\alpha}} \right\rangle : \forall x \in X \right\}
	\label{eq:3}
\end{equation} 

\section{Intuitionistic fuzzification of image}\label{Intfuzz}
Intuitionistic fuzzification of image is the process of transforming an image from crisp quantity to intuitionistic fuzzy form that comprise the uncertainty arise due to vagueness. Here, the image data is transformed in IFS data, that is in term of membership and non-membership degree using negation functions and then computed the hesitation degree. Mathematically, an image $B = \left\{ x_j \right\}_{j=1}^{N}$ having $N$ pixels, expressed as:

\begin{equation}
	B^{\text{IFS}} = \left\{ x_j^{\text{IFS}} \right\}_{j=1}^{N}
	\label{eq:4}
\end{equation}
where image data $(x_j)$ represented as $x_j^{\text{IFS}} = \left( \mu_B(x_j), \nu_B(x_j), \pi_B(x_j) \right)$  and $ \mu_B(x_j), \nu_B(x_j),$ and $\pi_B(x_j)$ represent the membership, non-membership and hesitation degree respectively of data $(x_j)$. The various membership function is used to compute the membership degree:

\begin{enumerate}
	\item 	The membership degree $\mu_B(x_j)$ of the data $x_j$ is computed as the normalized function, that is, convert the image data in the interval [0,1], defined as:  
	\begin{equation}
		\mu_B(x_j) = \frac{x_j - (x_j)_{\text{min}}}{(x_j)_{\text{max}} - (x_j)_{\text{min}}}
		\label{eq:5}
	\end{equation} 
	here $(x_j)_{\text{min}} \quad \text{and} \quad (x_j)_{\text{max}}$ represents the minimum and maximum values of data point $x_j$.         
	\item 	The Gaussian membership function is defined as $\mu_B(x_j) = \exp\left( - \frac{(x_j - c)^2}{2 \sigma^2} \right)$, where $c$ is the center of the Gaussian distribution and $\sigma$ is the standard deviation, which determines the spread or width of the Gaussian curve.
	
	\item 	The Sigmoid membership function is defined as $\mu_B(x_j) = \frac{1}{1 + \exp\left(-a(x_j - c)\right)}$, where $c$ is a parameter that determines the midpoint of the curve and $b$ is a parameter that determines the slope of the curve.
\end{enumerate}

\section{Proposed approach}\label{Pro}
The problem of uncertainty in feature representation using image segmentation methods like U-Net and U-Net++, which most prominently used for the image segmentation, is challenging issues to handle the uncertainty. In this section, we describe the framework to integrate the intuitionistic fuzzy set theory with U-Net and U-Net++ architecture. The details are described in the following subsections.

\subsection{Intuitionistic representation of data in triplet form }\label{tripletform}

\begin{figure}[htbp]
	\centering
	
	\begin{subfigure}[b]{0.15\textwidth}
		\centering
		\includegraphics[width=\textwidth]{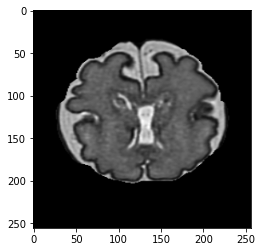}
		\caption{}
	\end{subfigure}
	\hfill
	\begin{subfigure}[b]{0.15\textwidth}
		\centering
		\includegraphics[width=\textwidth]{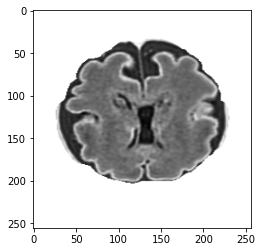}
		\caption{}
	\end{subfigure}
	\hfill
	\begin{subfigure}[b]{0.15\textwidth}
		\centering
		\includegraphics[width=\textwidth]{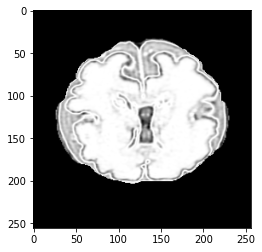}
		\caption{}
	\end{subfigure}
	
	\caption{Intuitionistic fuzzy data in form of (a): membership, (b): non-membership and (c): hesitation degree. The image corresponding to the membership degree shows a higher degree of belonging, while the image corresponding to non-membership degree highlight the respective the degree of non-belongingness, observable through the image contrast value. The image corresponding to the hesitation degree highlights the boundary region, capturing the uncertainty.}
	\label{fig:fign1}
\end{figure}

\begin{figure}[htbp]
	\centering
	
	\begin{subfigure}[b]{0.15\textwidth}
		\centering
		\includegraphics[width=\textwidth]{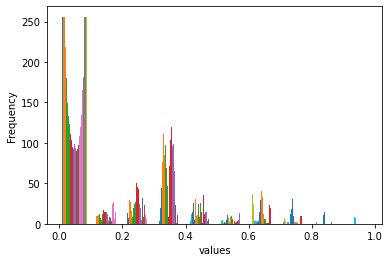}
		\caption{}
	\end{subfigure}
	\hfill
	\begin{subfigure}[b]{0.15\textwidth}
		\centering
		\includegraphics[width=\textwidth]{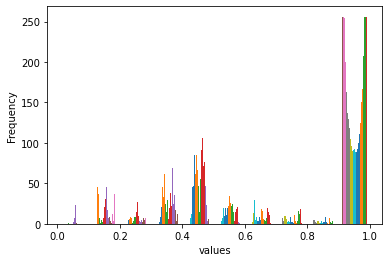}
		\caption{}
	\end{subfigure}
	\hfill
	\begin{subfigure}[b]{0.15\textwidth}
		\centering
		\includegraphics[width=\textwidth]{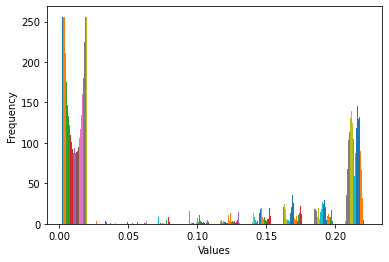}
		\caption{}
	\end{subfigure}
	
	\caption{Gray-level histogram corresponding to (a): membership degree image (b): non-membership image and (c) hesitation image.}
	\label{fig:fign2}
\end{figure}

The raw image is transformed into an intuitionistic fuzzy image $x_j^{\text{IFS}} = \left( \mu_B(x_j), \nu_B(x_j), \pi_B(x_j) \right)$  \eqref{eq:4} to capture the uncertainty in transition region, and further helping to classify each pixel into the desired class. The input image is represented as $\left( \mu_B(x_j), \nu_B(x_j), \pi_B(x_j) \right)$ in term of membership, non-membership and hesitation degree. The intuitionistic representation of image exploited the qualitative information $\left( \mu_B(x_j), \nu_B(x_j), \pi_B(x_j) \right)$ in comparison to classical representation of an image. The features in an intuitionistic way using intuitionistic fuzzy sets, facilitating the classification of data into relevant classes. This process is illustrated in Fig. \ref{fig:fign1}, and the corresponding histogram of the image is depicted in Fig. \ref{fig:fign2}. The image corresponding to the hesitation degree visualizes the boundary region and provides more information to extract relevant features, creating intuition within the intuitionistic fuzzy set environment. This transformation captures the uncertainty in the transition region of the image along with the relevant features, enabling more accurate pixel classification. The intuitionistic fuzzy representation leverages qualitative information about image features, encapsulating uncertainty through these three degrees.

The uncertainty of image feature values is represented by values between zero to one. A large value of hesitation indicates the more uncertainty like partial volume effect in medical images. The histogram of corresponding image is shown in Fig.\ref{fig:fign2} The image corresponding to membership and non-membership denoted with various values shows the degree of belonging and non-belonging into the relevant classes. The zero-hesitation value shows the no uncertainty and strongly belong to the relevant classes. 

\begin{figure*}[h!]
	\centering
	\includegraphics[width=0.85\textwidth]{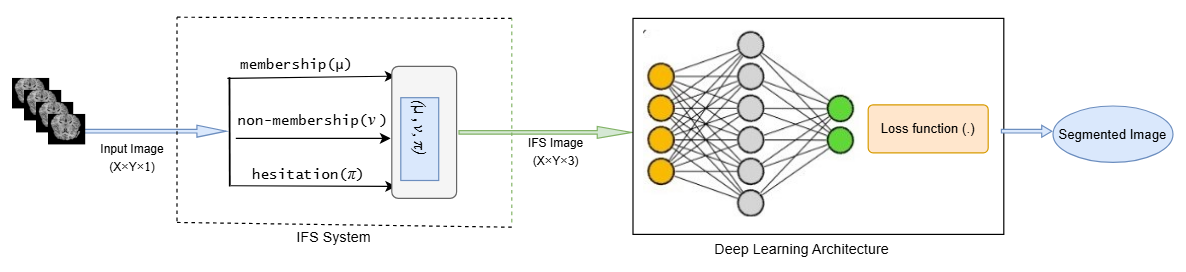}
	\caption{An overview of the proposed method in the deep learning architecture. The schematic overview of the training process of intuitionistic fuzzy set (IFS) system representation and deep learning architecture for segmentation of brain image. In this process, IFS system converted the input image into the triplet (intuitionistic fuzzy set) form that is membership $(\mu)$, non-membership $(\nu)$ and hesitation degree $(\pi)$. The uncertainty features in the image exploited the qualitative information using intuitionistic fuzzy logic in comparison to classical representation of an image. It utilizes the uncertainty during the training process.}
	\label{fig:fign3}
\end{figure*}

This representation of raw image $\{x_j\}_{j=1}^N$ into an intuitionistic fuzzy image $x_j^{\mathrm{IFS}} = \left( \mu_B(x_j), \nu_B(x_j), \pi_B(x_j) \right)$, which capture the uncertainty using intuitionistic fuzzy set, known as intuitionistic fuzzy set (IFS) system, depicted in Fig. \ref{fig:fign3}. The IFS system converts the image into triplet form that is used the input for deep learning architectures. The intuitionistic image further passes as input to deep learning architecture, shown in Fig. \ref{fig:fign3}.  

\subsection{Incorporation of intuitionistic fuzzy logic }\label{IFlogic}
IFS system integrated with the deep learning architectures for image segmentation like U-Net and U-Net during training process to enhance image segmentation accuracy, particularly in uncertain region, which are described in the following subsections.

\subsubsection{Intuitionistic fuzzy set with U-Net ( \text{IFS\_U-Net})  }\label{IFSUNet}
A variant of the convolutional neural network architecture, named U-Net, is designed by Ronneberger et al. \cite{Ronneberger} for biomedical image segmentation. It has U shape with a contractive path (down-sampling) to capture relevant features and an expanding path (up-sampling) to enable precise localization in the architecture. The down-sampling path comprises convolution and max pooling, while the up-sampling part restores spatial information pixel by pixel and combines features from the contracting part to preserve information. The U-Net effectively segments biomedical images with high accuracy and is widely used in medical image analysis. However, the existing U-Net, which uses deep learning concepts, is unable to identify uncertainties due to boundary regions and poor contrast.

\begin{figure*}[h!]
	\centering
	\includegraphics[width=0.7\textwidth]{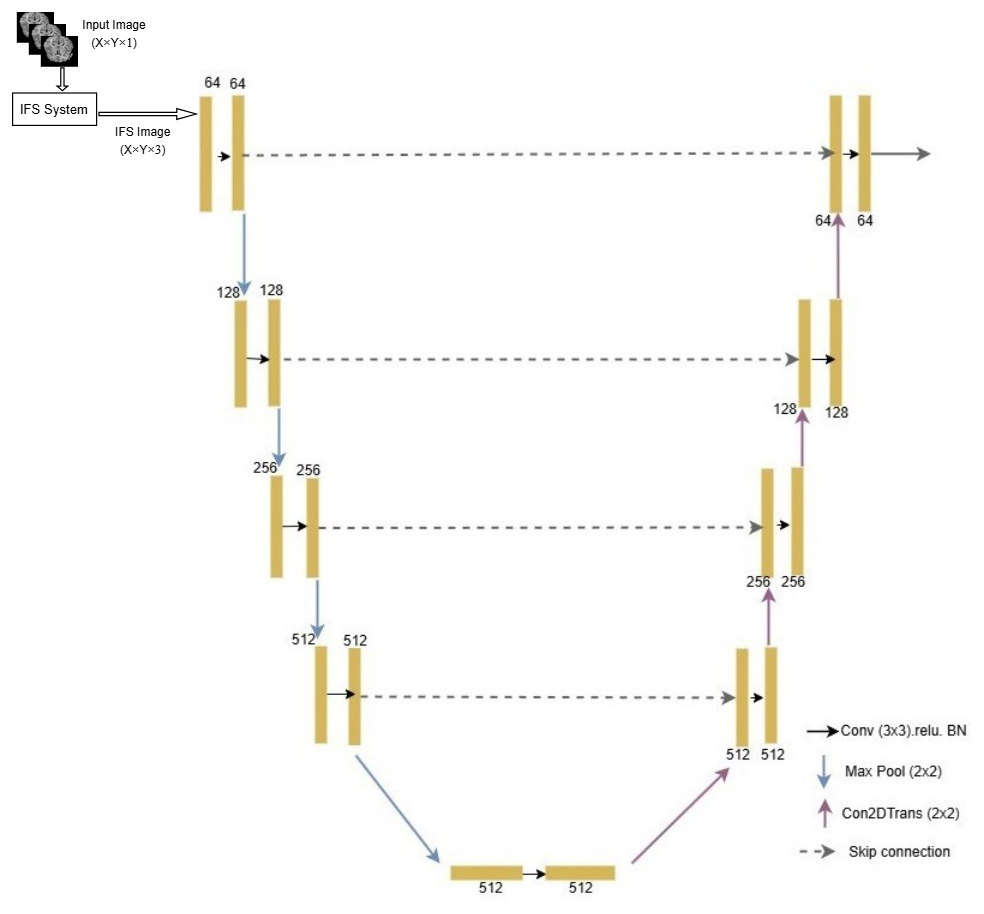}
	\caption{\text{IFS\_U-Net} framework with intuitionistic fuzzy logic systems. It accepts the input image in term of intuitionistic fuzzification of data and utilizes the uncertainty during the training process. And as a result, it able to segment the brain tissues in more accurate.}
	\label{fig:fign4}
\end{figure*}

In order to incorporate the intuitionistic intuition to deal the uncertainty, we incorporate IFS system (shown in Fig. \ref{fig:fign3}) in U-Net and designed framework, named as \text{IFS\_U-Net}, shown in Fig. \ref{fig:fign4}. In this novel framework, firstly feature of input data is encoded in the intuitionistic fuzzy domain with converting the image in intuitionistic fuzzy representation terms $x_j^{\text{IFS}} = \left( \mu_B(x_j), \nu_B(x_j), \pi_B(x_j) \right)$ and then processed for deep learning training.  Each pixel of the image is converted into qualitative features through intuitionistic fuzzy encoding. In brain images, brain tissues have varying intensities values. The intuitionistic fuzzy representation of the data extracts features across different intensity levels. Among these three data features, the membership degree extracts the belongingness feature, while the non-membership degree extracts the non-belonging feature from the data. The hesitation part captures the uncertainty features across various boundaries in the brain image with varying intensity values. This includes transforming the data into a three-way representation in form of a vector consisting of membership, non-membership, and hesitation degrees. In this triple vector space, intuitionistic logic is used to deal the uncertainty from different tissues. 

The down-sampling part accepts the input data in the form of membership, non-membership, and hesitation degree, and extracts the uncertain relevant features using intuitionistic fuzzy logic. This approach helps to handle boundary pixels which shows the vagueness, as the data is interpreted through intuitionistic fuzzy logic. Membership value shows that pixels belong to a similar group, non-membership degree shows that pixels belong to a different group, and hesitation degree determines the boundary pixels' assignment to a particular group, aiding in deciding the classification of pixels. The up-sampling part examines the location of the pixels, and intuitionistic logic assists in decision-making. Determining the boundary region in brain images is particularly challenging when the fuzzy values for two classes are very close. Intuitionistic fuzzy logic improves decision-making and classifies the pixels into the desired classes are more effectively.

\subsubsection{Intuitionistic fuzzy logic with U-Net++ ( \text{IFS\_U-Net++})  }\label{IFSUNet}

U-Net++ is an advanced segmentation architecture introduced by Zhou et al. \cite{Zhou2018,Zhou2019} to enhance the accuracy of medical image segmentation by incorporating a series of nested and dense skip connections. This novel approach is designed to bridge the semantic gap between encoder and decoder feature maps, ensuring a more refined extraction of object boundaries and intricate structures in medical images. The fundamental principle behind U-Net++ is the progressive refinement of high-resolution feature maps within the encoder network before their integration with the semantically enriched feature maps of the decoder. With this, model becomes more capable in capturing the finer details of foreground objects, leading to more precise segmentation outcomes. Unlike the traditional U-Net, where simple skip connections directly transfer high-resolution features from the encoder to the decoder, resulting in the fusion of feature maps with varying semantic meanings, U-Net++ mitigates this issue through its redesigned skip pathways. U-Net++ consists of an encoder sub-network (backbone) followed by a decoder sub-network. The primary distinction between U-Net++ and U-Net lies in its three key enhancements:

\begin{itemize}
	\item Convolutional Layers in Skip Connections (Green Pathways): Instead of direct feature transfer, U-Net++ introduces convolutional layers within skip connections, ensuring that feature maps shared between the encoder and decoder remain semantically aligned. 
	\item Dense Skip Connections (Blue Pathways): U-Net++ leverages dense connections along the skip pathways, which facilitate smoother gradient flow and improve feature reuse, ultimately leading to better convergence and performance. 
	\item Deep Supervision (Red Pathways): It employs deep supervision by integrating auxiliary classifiers at multiple decoder levels. This ensures that the model optimizes feature maps at various depths, leading to a more effective training process and improved segmentation accuracy. 
\end{itemize}

\begin{figure*}[h!]
	\centering
	\includegraphics[width=0.7\textwidth]{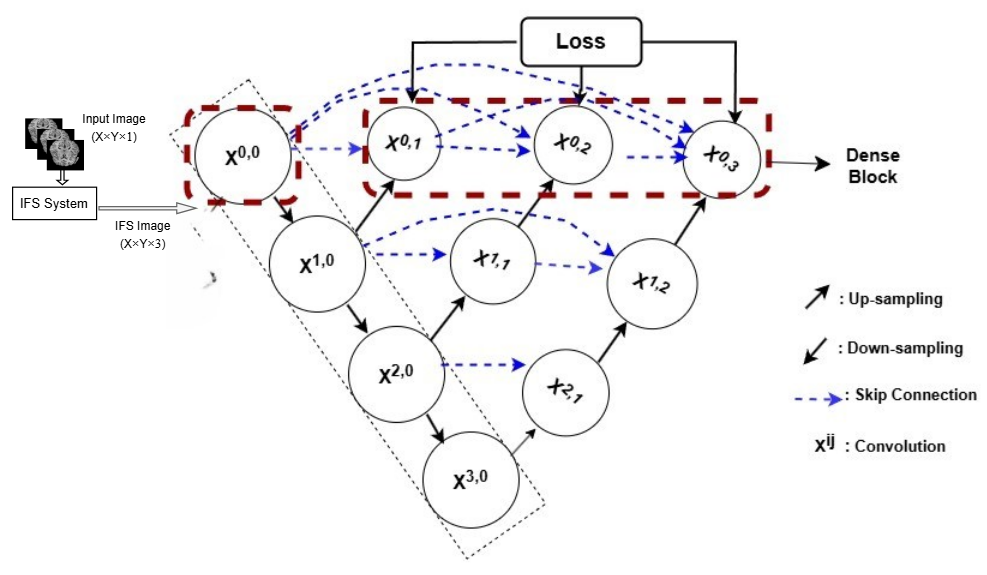}
	\caption{ Framework for \text{IFS\_U-Net++} with intuitionistic fuzzy logic systems, where the input image accepted in term of intuitionistic fuzzification of data and utilizes the uncertainty. The input image $(X \times Y \times 1)$ converted in form of $(X \times Y \times 3)$) as the input for \text{IFS\_U-Net++} architecture}
	\label{fig:fign5}
\end{figure*}

With incorporating these enhancements, U-Net++ effectively addresses the limitations of U-Net architecture, making it particularly well-suited for applications requiring high-precision segmentation, such as medical image analysis. 

To effectively capture useful information in uncertain scenario, we integrate the IFS system (shown in Fig.\ref{fig:fign3} with U-Net++, and designed a novel framework, named as \text{IFS\_U-Net++}, depicted in Fig. \ref{fig:fign5}. In this system, input data features are encoded within an intuitionistic fuzzy domain, where each image pixel is transformed into an intuitionistic fuzzy representation: $x_j^{\text{IFS}} = \left( \mu_B(x_j), \nu_B(x_j), \pi_B(x_j) \right)$. The membership degree captures the extent to which a pixel belongs to a particular class, while the non-membership degree indicates exclusion from that class. The hesitation degree quantifies the uncertainty present in boundary regions where intensity differences are subtle. Intuitionistic fuzzy logic helps to deal the uncertainty in \text{IFS\_U-Net++} by transforming the data into a three-component representation, facilitating feature extraction from different brain tissues. This approach effectively handles boundary pixels, which often exhibit vagueness. By incorporating intuitionistic fuzzy logic, the model achieves more precise pixel classification and enhances decision-making.

\section{Experiment}\label{Exp}
\subsection{Dataset}
This study utilizes two publicly available MRI brain datasets, named IBSR (Internet Brain Segmentation Repository) and OASIS (Open Access Series of Imaging Studies), which play a crucial role in neuroimaging research.\\ 
\textit{IBSR brain data:} The IBSR (https://www.nitrc.org/projects/ibsr/) brain data used for evaluating segmentation techniques by offering manually segmented brain data. It includes 20 normal T1-weighted MRI brain scans along with corresponding expert-annotated segmentation data. This dataset is well-labelled, detailing various brain structures and tissues, which is crucial for training and validating segmentation algorithms. The annotated labels act as ground truth image, enabling the assessment of automated segmentation methods. The dataset is provided by Center for Morphometric Analysis at Massachusetts General Hospital.\\
\textit{OASIS brain data:} It comprises a diverse set of MRI scans from both healthy individuals and patients with cognitive impairments, accompanied by extensive clinical metadata. The OASIS (http://www.oasis-brains.org/) brain data offers publicly accessible brain imaging data for research and analysis. The dataset includes 416 subjects aged between 18 and 96, with each subject having 3 or 4 T1-weighted MRI scans taken in a single session. Non-brain tissues are removed using the Brain Extraction Tool. Additionally, the dataset is well-annotated with segmentation labels that identify different brain structures and tissues.

\subsection{Baseline and implementation}

We implement, the novel farmwork viz \text{IFS\_U-Net} and \text{IFS\_U-Net++}, where image processed into an intuitionistic fuzzy representation form $x_j^{\text{IFS}} = \left( \mu_B(x_j), \nu_B(x_j), \pi_B(x_j) \right)$ using the Sugano’s and Yager negation function. The input data in\text{IFS\_U-Net} and \text{IFS\_U-Net++} accepted as data in triplet of vector to extract the relevant information in uncertainty scenario. To test the segmentation performance of proposed framework: \text{IFS\_U-Net} and \text{IFS\_U-Net++}, we tested these architectures on MRI brain image and compared their performance with U-Net and U-Net++ and further study the ablation part. We use categorical cross-entropy optimization loss function in training the model. The ReLU activation function and Adam optimizer are utilized in U-Net, U-Net++, \text{IFS\_U-Net} and \text{IFS\_U-Net++}. Additionally, batch normalization and dropout layers are applied after each convolutional layer in U-Net++ and \text{IFS\_U-Net++} to mitigate overfitting, while batch normalization layers are integrated within U-Net framework. A kernel size of $3 \times 3$ is used, with details of the neuron configuration depicted in Fig. \ref{fig:fign4} and Fig. \ref{fig:fign5}. The experiments are conducted using Keras with a TensorFlow backend. To quantitative comparison for segmentation performance of various architectures, we use various evaluation metrics like Accuracy $(AC)$, Dice coefficient $(DC)$, and Intersection over Union $(IoU)$ in this study. We take the number of epochs 100 for IBSR data and used early stoppage for OASIS data. In entire experiment, batch size 2 is taken for both brain data. The IBSR and OASIS brain data are split into 80:20 ratio, where 80\% used for training and model and 20\% data for the testing. 

\section{Results and discussion}\label{Resdis}
\begin{figure*}  
	\centering
	\begin{subfigure}[t]{0.45\textwidth}
		\centering
		\includegraphics[width=\textwidth]{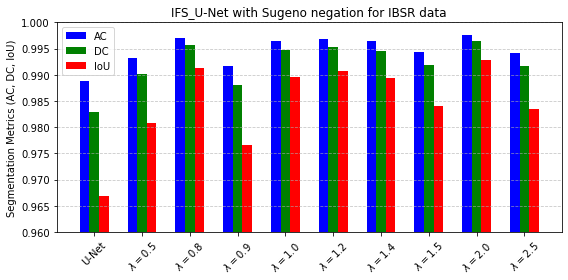}
	\end{subfigure}
	\hfill
	\begin{subfigure}[t]{0.45\textwidth}
		\centering
		\includegraphics[width=\textwidth]{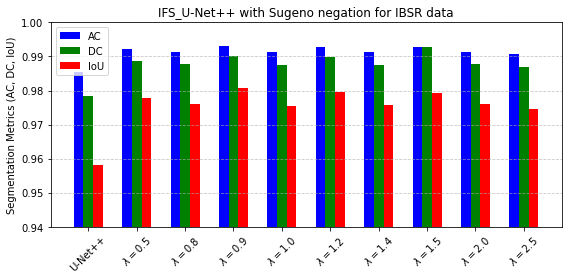}
	\end{subfigure}
	
	\vspace{2em}
	
	\begin{subfigure}[t]{0.45\textwidth}
		\centering
		\includegraphics[width=\textwidth]{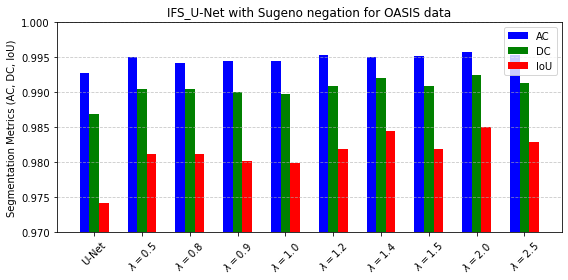}
	\end{subfigure}
	\hfill
	\begin{subfigure}[t]{0.45\textwidth}
		\centering
		\includegraphics[width=\textwidth]{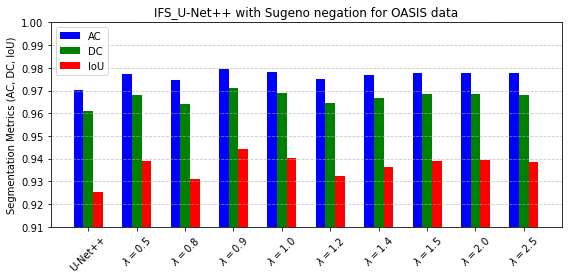}
	\end{subfigure}
	
	\vspace{2em}
	
	\caption{Ablation study for segmentation performance of U-Net, U-Net++ and proposed framework: \text{IFS\_U-Net}, \text{IFS\_U-Net++} with Sugneo negation function in term of $(AC,DC,IoU)$ for IBSR and OASIS data with various value of $\lambda$ in Sugeno negation function. The proposed systems: \text{IFS\_U-Net}, \text{IFS\_U-Net++} with Sugneo negation function for all considered value of $\lambda$ provide better consistent segmentation results in comparison to U-Net, U-Net++ utilizes the vague and qualitative information. }
	\label{fig:fign6}
\end{figure*}

Ablation study of the proposed system (\text{IFS\_U-Net}, \text{IFS\_U-Net++}) for MRI segmentation in the association of the intuitionistic fuzzy logic with Sugeno and Yager negation function, and deep learning architectures: U-Net, U-Net++ with categorical cross-entropy loss function has been carried on IBSR and OASIS brain dataset. Firstly, we tested U-Net, U-Net++, \text{IFS\_U-Net}, \text{IFS\_U-Net++} on IBSR data. All 20 cases from the IBSR dataset taken in this study along with labelled data. In OASIS data, we take 704 brain 2-D images to train the model. The corresponding labelled image is used to train the supervised model. The 20\% images are considered for the testing and remaining 80\% images are used for the training. Table \ref{tab:tabn1} , comprises the quantitative segmentation results and compare the efficacy of segmentation performance in term of $AC$, $DC$, and $IoU$ for IBSR and OASIS data for U-Net and proposed \text{IFS\_U-Net} framework with Sugeno negation function with various value of $\lambda$ in Sugeno negation functions. The intuitionistic fuzzy logic with Sugeno and Yager negation function handle uncertainty between different tissues. From, Table \ref{tab:tabn1}, it is evident that proposed framework: \text{IFS\_U-Net}, \text{IFS\_U-Net++} with Sugeno negation function perform better in comparison to U-Net for IBSR and OASIS data. The \text{IFS\_U-Net} with $\lambda=2.0$ give the best $AC (=0.9976),DC(=0.9964)$ and $IoU(=0.9929)$ for IBSR data and  $AC(=0.9958),DC(=0.9924)$ and $IoU(=0.9850)$ for OASIS data. The quantitative segmentation performance in term of $AC,DC$ and $IoU$ are also shown in the Fig. \ref{fig:fign6}(a) and Fig. \ref{fig:fign6}(c) using bar diagram, which clearly demonstrate the efficacy of \text{IFS\_U-Net}.

\begin{table*}[htbp]
	\centering
	\caption{Quantitative comparison of segmentation performance for U-Net, and \text{IFS\_U-Net} with intuitionistic fuzzy logic with Sugeno negation function along with various value of $\lambda$ for IBSR and OASIS data. We take the number of epochs 100 for IBSR data and use early stoppage for OASIS data and take the Batch size 2 for both brain data. The quantitative segmentation performances presented in term of accuracy $(AC)$, Dice coefficient $(DC)$ and Intersection over Union $(IoU)$. Best respective results are marked in \textbf{bold}.}
	
	\label{tab:tabn1}
	\begin{tabular}{llcccccccccc}
		\toprule
		
		\text{Data} & \text{Metrics} & \text{U-Net} & \multicolumn{9}{c}{\text{IFS\_U-Net with Sugeno negation function}} \\
		\cline{4-12}
		& & & $\lambda=0.5$ & $\lambda=0.8$ & $\lambda=0.9$ & $\lambda=1.0$ & $\lambda=1.2$ & $\lambda=1.4$ & $\lambda=1.5$ & $\lambda=2.0$ & $\lambda=2.5$ \\
		
		\midrule
		
		\multirow{3}{*}{IBSR} 
		& AC  & 0.9888 & 0.9932 & 0.9971 & 0.9916 & 0.9964 & 0.9968 & 0.9964 & 0.9943 & 0.9976 & 0.9942 \\
		& DC  & 0.9829 & 0.9901 & 0.9956 & 0.9880 & 0.9947 & 0.9953 & 0.9946 & 0.9918 & 0.9964 & 0.9916 \\
		& IoU & 0.9668 & 0.9807 & 0.9913 & 0.9765 & 0.9895 & 0.9908 & 0.9893 & 0.9840 & 0.9929 & 0.9835 \\
		
		\midrule
		
		\multirow{3}{*}{OASIS} 
		& AC  & 0.9928 & 0.9950 & 0.9942 & 0.9945 & 0.9945 & 0.9954 & 0.9950 & 0.9952 & 0.9958 & 0.9954 \\
		& DC  & 0.9869 & 0.9905 & 0.9904 & 0.9900 & 0.9898 & 0.9909 & 0.9921 & 0.9909 & 0.9924 & 0.9913 \\
		& IoU & 0.9742 & 0.9812 & 0.9811 & 0.9802 & 0.9799 & 0.9819 & 0.9844 & 0.9819 & 0.9850 & 0.9829 \\
		
		\bottomrule
		
	\end{tabular}
\end{table*}

Further, we evaluate U-Net++ and \text{IFS\_U-Net++} with Sugeno negation function, with various value of $\lambda$ for IBSR and OASIS data. The quantitative segmentation results in term of $AC,DC$ and $IoU$ are included in Table \ref{tab:tabn2}. Intuitionistic fuzzy logic, utilizing Sugeno negation function with various value of $\lambda$ handle the uncertainty and provide better segmentation performance on IBSR and OASIS data, as presented in Table \ref{tab:tabn2}. Also from the bar diagram, represented in Fig. \ref{fig:fign6} (b) and Fig. \ref{fig:fign6}(d), \text{IFS\_U-Net++} with Sugeno negation function provide consistent superior performance relative to U-Net for IBSR and OASIS datasets. The proposed \text{IFS\_U-Net++} framework at $\lambda(=0.9)$ give the best segmentation value in term of $AC (=0.9931),DC(=0.9901)$ and $IoU(=0.9806)$ for IBSR data, and $AC (=0.9795),DC(=0.9710)$ and $IoU(=0.9441)$ for OASIS data, marked with bold.

\begin{table*}[htbp]
	\centering
	\caption{Quantitative comparison of segmentation performance for U-Net++ and \text{IFS\_U-Net++} with intuitionistic fuzzy logic with Sugeno negation function, with various value of $\lambda$ for IBSR and OASIS data. We take the number of epochs 100 for IBSR data and use early stoppage for OASIS data and Batch size 2 for both brain data.  The segmentation performances represented in term of accuracy $(AC)$, Dice coefficient $(DC)$ and Intersection over Union $(IoU)$. Best respective results are marked in \textbf{bold}.}
	
	\label{tab:tabn2}
	\begin{tabular}{l l c c c c c c c c c c}
		\toprule
		\text{Data} & \text{Metrics} & \text{U-Net++} & \multicolumn{9}{c}{\text{IFS\_U-Net++ with Sugeno negation function}} \\
		\cline{4-12}
		& & & $\lambda=0.5$ & $\lambda=0.8$ & $\lambda=0.9$ & $\lambda=1.0$ & $\lambda=1.2$ & $\lambda=1.4$ & $\lambda=1.5$ & $\lambda=2.0$ & $\lambda=2.5$ \\
		\midrule
		\multirow{3}{*}{IBSR} 
		& AC  & 0.9853 & 0.9922 & 0.9914 & 0.9931 & 0.9912 & 0.9928 & 0.9912 & 0.9927 & 0.9914 & 0.9908 \\
		& DC  & 0.9785 & 0.9887 & 0.9877 & 0.9901 & 0.9874 & 0.9897 & 0.9875 & 0.9927 & 0.9877 & 0.9869 \\
		& IoU & 0.9580 & 0.9779 & 0.9761 & 0.9806 & 0.9754 & 0.9797 & 0.9756 & 0.9793 & 0.9760 & 0.9746 \\
		\midrule
		\multirow{3}{*}{OASIS} 
		& AC  & 0.9704 & 0.9774 & 0.9745 & 0.9795 & 0.9781 & 0.9751 & 0.9769 & 0.9775 & 0.9778 & 0.9778 \\
		& DC  & 0.9609 & 0.9682 & 0.9640 & 0.9710 & 0.9689 & 0.9647 & 0.9669 & 0.9683 & 0.9685 & 0.9681 \\
		& IoU & 0.9254 & 0.9388 & 0.9311 & 0.9441 & 0.9401 & 0.9324 & 0.9365 & 0.9391 & 0.9394 & 0.9386 \\
		\bottomrule
	\end{tabular}
\end{table*}

\begin{figure*}  
	\centering
	\begin{subfigure}[t]{0.45\textwidth}
		\centering
		\includegraphics[width=\textwidth]{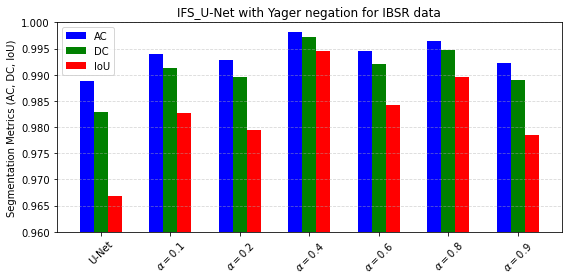}
	\end{subfigure}
	\hfill
	\begin{subfigure}[t]{0.45\textwidth}
		\centering
		\includegraphics[width=\textwidth]{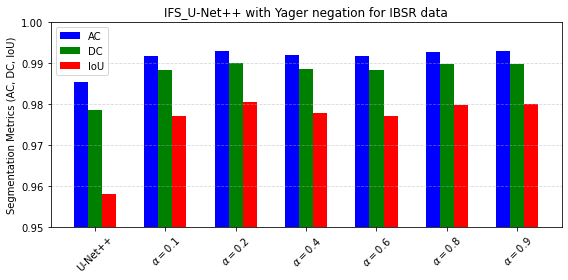}
	\end{subfigure}
	
	\vspace{2em}
	
	\begin{subfigure}[t]{0.45\textwidth}
		\centering
		\includegraphics[width=\textwidth]{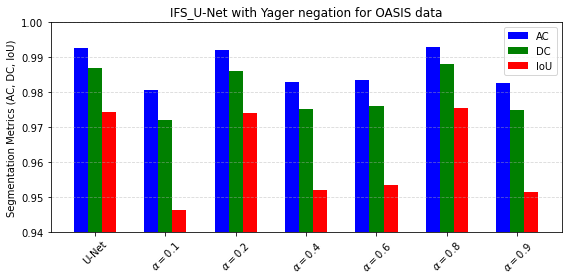}
	\end{subfigure}
	\hfill
	\begin{subfigure}[t]{0.45\textwidth}
		\centering
		\includegraphics[width=\textwidth]{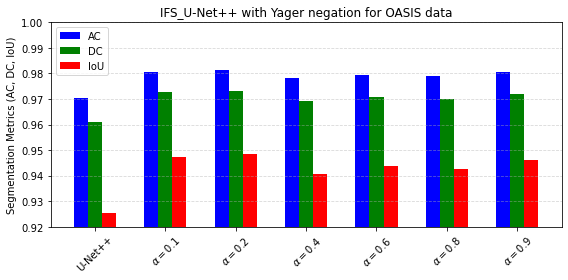}
	\end{subfigure}
	
	\vspace{2em}
	
	\caption{Ablation study of segmentation performance $(AC,DC,IoU)$ for U-Net, U-Net++ and \text{IFS\_U-Net} , \text{IFS\_U-Net++} with Yager negation function for IBSR and OASIS data with various value of $\alpha$. The proposed framework: \text{IFS\_U-Net}, \text{IFS\_U-Net++} with Yager negation function for all considered value of $\alpha$  provide better segmentation results in comparison to U-Net, U-Net++ with utilizing the vague and qualitative information for IBSR data. On OASIS data,  \text{IFS\_U-Net}, results are equal or less than U-Net architecture, however the  \text{IFS\_U-Net++} results is better in comparison to U-Net++. }
	\label{fig:fign7}
\end{figure*}

To check the efficacy of \text{IFS\_U-Net} with Yager negation function, we trained U-Net and \text{IFS\_U-Net} with Yager negation function, with various value of $\alpha$  for IBSR and OASIS data. The quantitative segmentation results in term of $AC, DC$, and $IoU$ are presented in Table \ref{tab:tabn3} for IBSR and OASIS data. Intuitionistic fuzzy logic, utilizing Yager negation function with various value of $\alpha$ handle the uncertainty and partial volume effect and make better segmentation performance on IBSR and OASIS data, as presented in the Table \ref{tab:tabn3}. The \text{IFS\_U-Net} with Yager negation function provide better results for all value of $\alpha$ on IBSR data. However, it provides the equal performance at $\alpha(=0.4)$ on OASIS data and other segmentation result on OASIS data is not good. It provided consistent performance for IBSR data. The quantitative comparison is also shown with bar diagram represented in Fig. \ref{fig:fign7}. Also, from the bar diagram (Fig. \ref{fig:fign7} (a) and Fig. \ref{fig:fign7}(c)), proposed \text{IFS\_U-Net} with the Yager negation function provide superior performance relative to U-Net on IBSR. For OASIS data, its performance is equal or less than the original architecture. \text{IFS\_U-Net} at $\alpha=0.4$ give the highest segmentation value in term of $AC(=0.9982),DC(=0.9972)$ and $IoU(=0.9945)$ for IBSR data, and $AC (=0.9930),DC(=0.9882)$ and $IoU(=0.9744)$ for OASIS data.

\begin{table*}[htbp]
	\centering
	\caption{The quantitative comparison of segmentation performance with various value of $\alpha$  for IBSR and OASIS data for U-Net, and \text{IFS\_U-Net} with intuitionistic fuzzy logic using Yager negation function to handle the uncertainty. We take the number of epochs 100 for IBSR data and use early stoppage for OASIS data and Batch size 2 for both brain data. The segmentation performances evaluated in term of $AC, DC$, and $IoU$.}
	\label{tab:tabn3}
	\begin{tabular}{c c c c c c c c c}
		\toprule
		\text{Data} & \text{Metrics} & \text{U-Net}& \multicolumn{6}{c}{\text{IFS\_U-Net with Yager negation function}} \\
		\cline{4-9}
		
		& & & $\alpha=0.1$ & $\alpha=0.2$ & $\alpha=0.4$ & $\alpha=0.6$ & $\alpha=0.8$ & $\alpha=0.9$ \\
		
		\midrule
		\multirow{3}{*}{IBSR} 
		& AC  & 0.9888 & 0.9939 & 0.9928 & 0.9982 & 0.9945 & 0.9964 & 0.9923 \\
		& DC  & 0.9829 & 0.9912 & 0.9895 & 0.9972 & 0.9920 & 0.9947 & 0.9890 \\
		& IoU & 0.9668 & 0.9827 & 0.9794 & 0.9945 & 0.9843 & 0.9895 & 0.9784 \\
		\midrule
		\multirow{3}{*}{OASIS} 
		& AC  & 0.9928 & 0.9806 & 0.9921 & 0.9930 & 0.9835 & 0.9828 & 0.9827 \\
		& DC  & 0.9869 & 0.9721 & 0.9861 & 0.9882 & 0.9760 & 0.9752 & 0.9750 \\
		& IoU & 0.9742 & 0.9461 & 0.9740 & 0.9755 & 0.9533 & 0.9520 & 0.9515 \\
		\bottomrule
	\end{tabular}
\end{table*}

Further, ablation study of U-Net++ and \text{IFS\_U-Net++} with Yager negation function with various value of $\alpha$ tested on IBSR and OASIS data and computed their quantitative segmentation results in term of $AC,DC$, and $IoU$, are presented in Table \ref{tab:tabn4}. Intuitionistic fuzzy logic, using Yager negation function handle the uncertainty and make better segmentation performance for IBSR and OASIS data with handling the vagueness in brain tissues. Also, bar diagram of segmentation performance $(AC,DC, and IoU)$ are represented in Fig. \ref{fig:fign7}. From the bar diagram in Fig. \ref{fig:fign7} (b) and Fig. \ref{fig:fign7}(d), \text{IFS\_U-Net++} with Yager negation function for various value of parameter $\alpha$ provide consistent superior segmentation performance relative to U-Net++ architecture for IBSR and OASIS datasets. The \text{IFS\_U-Net++} with $\alpha=0.2$ give the highest segmentation value of $AC (=0.9931),DC(=0.9901)$ and $IoU(=0.9806)$ for IBSR data and $AC (=0.9812),DC(=0.9733)$ and $IoU(=0.9483)$ for OASIS data.

\begin{table*}[htbp]
	\centering
	\caption{The quantitative segmentation performance $(AC, DC,$ and $IoU)$ for U-Net++ and suggested \text{IFS\_U-Net++} with intuitionistic fuzzy logic using Yager negation function with various value of $\alpha$  for IBSR and OASIS data. We take the number of epochs 100 for IBSR data and use early stoppage for OASIS data and take the Batch size 2 for both brain data.}
	\label{tab:tabn4}
	\begin{tabular}{c c c c c c c c c}
		\toprule
		\text{Data} & \text{Metrics} & \text{U-Net++}& \multicolumn{6}{c}{\text{IFS\_U-Net++ with Yager negation function}} \\
		\cline{4-9}
		
		& & & $\alpha=0.1$ & $\alpha=0.2$ & $\alpha=0.4$ & $\alpha=0.6$ & $\alpha=0.8$ & $\alpha=0.9$ \\
		
		\midrule
		\multirow{3}{*}{IBSR} 
		& AC  & 0.9853 & 0.9919 & 0.9931 & 0.9921 & 0.9918 & 0.9928 & 0.9929 \\
		& DC  & 0.9785 & 0.9883 & 0.9901 & 0.9887 & 0.9883 & 0.9897 & 0.9898 \\
		& IoU & 0.9580 & 0.9772 & 0.9805 & 0.9778 & 0.9772 & 0.9797 & 0.9801 \\
		\midrule
		\multirow{3}{*}{OASIS} 
		& AC  & 0.9704 & 0.9805 & 0.9812 & 0.9782 & 0.9793 & 0.9792 & 0.9804 \\
		& DC  & 0.9609 & 0.9727 & 0.9733 & 0.9691 & 0.9709 & 0.9702 & 0.9721 \\
		& IoU & 0.9254 & 0.9472 & 0.9483 & 0.9405 & 0.9439 & 0.9425 & 0.9460 \\
		\bottomrule
	\end{tabular}
\end{table*}

\begin{figure*}[h!]
	\centering
	\includegraphics[width=0.7\textwidth]{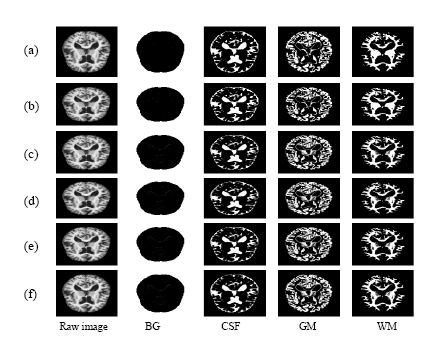}
	\caption{Visual representation of OASIS brain image segmentation of a particular slice (2D image) using deep learning architecture into different classes like BG (background), CSF, GM and WM. Figure shows image segmentation with original and intuitionistic fuzzy logics using Sugeno and Yager negation function, architectures: (a) U-Net, (b) \text{IFS\_U-Net} ($\lambda=2.0$), (c) \text{IFS\_U-Net} ($\alpha=0.4 $), (d) U-Net++, (e) \text{IFS\_U-Net++} ($\lambda=0.9$) (f) \text{IFS\_U-Net++} ($\alpha=0.4$). The deep learning architectures with intuitionistic fuzzy logic handle the partial volume effect and provide the better segmentation.  }
	
	\label{fig:fign8}
\end{figure*}

In this study, we conduct the ablation study and compute the quantitative result in term of metrices $(AC,DC,IoU)$. The qualitative segmentation results of tissues are represented in Fig. \ref{fig:fign8}, using the deep learning architectures, main objective of brain image segmentation in various tissues. In Fig. \ref{fig:fign8}, it shows the raw image and segmented tissues of brain image such as CSF, GM, WM and background part. It is tedious work to validate the visual brain image segmentation. Deep learning architectures are trained with brain data and then predicted the segmented image, which is represented in Fig. \ref{fig:fign8}. 

Overall, Tables, \ref{tab:tabn1}, \ref{tab:tabn2}, \ref{tab:tabn3}, and \ref{tab:tabn4}, provides strong evidence for effectiveness of \text{IFS\_U-Net} and \text{IFS\_U-Net++} on both brain data, except one case, with utilizes the uncertainty in training process. We check the effect of parameter of $\lambda$ in Sugeno negation function and $\alpha$ in Yager negation function, in construction of intuitionistic fuzzy data on both IBSR and OASIS brain data. Quantitative segmentation performance of \text{IFS\_U-Net} and \text{IFS\_U-Net++} varies for various value of $\lambda$ and $\alpha$, and promising the better result for all value in comparison to original architecture. We include the ablation studies on effect of intuitionistic fuzzy logic in U-Net and U-Net++ architectures, which unitizing the qualitative information. In contrast to U-Net and U-Net++ architectures, \text{IFS\_U-Net} and \text{IFS\_U-Net++} fully leverage the prediction segmentation with intuitionistic fuzzy set theory. 

\section{Conclusion }\label{con}

To handle the uncertainty in feature representation in a biomedical image segmentation process through deep learning architecture, in this work, we presented a novel framework, called \text{IFS\_U-Net} and \text{IFS\_U-Net++} with integrating the intuitionistic fuzzy logic. The input data in \text{IFS\_U-Net} and \text{IFS\_U-Net++} are represented in an intuitionistic fuzzy form to effectively handle the uncertainty between the region of interest and non-region of interest.  The \text{IFS\_U-Net} and \text{IFS\_U-Net++} utilize the uncertainty in the data image during the training process to enhance the segmentation performance. The experiment is conducted on IBSR and OASIS brain data, and quantitative segmentation performance is computed in term of AC, DC, and IoU for comparison and to validate the proposed novel system with intuitionistic fuzzy logic and U-Net \& U-Net++. The ablation study and experiment results show the proposed framework:\text{IFS\_U-Net} and \text{IFS\_U-Net++} provides consistent better result for IBSR and OASIS data for different value of parameters in Sugneo and Yager negation function. In the future, the integration of intuitionistic fuzzy logic in deep learning will enhance accuracy and reliability in handling uncertain environments.

\section*{Acknowledgment}
This research project is supported by the Innovative Research Grant (IRG), file No: IRG/MJPRU/DoR/2022/06, under the Directorate of Research, MJP Rohilkhand University, Bareilly, Uttar Pradesh, India. The financial assistance is greatly appreciated and has played a crucial role in the execution of this research.

\end{document}